\documentclass[11pt]{article}
\usepackage{moriond,epsfig}

\def\Journal#1#2#3#4{{#1} {\bf #2}, #3 (#4)}

\def\NIMA{{\em Nucl. Instrum. Methods} A}

\def\PLB{{\em Phys. Lett.}  B}

\def\APP{\em Astropart. Phys.}

\def\be{\begin{equation}}
\def\ee{\end{equation}}
\def\bea{\begin{eqnarray}}
\def\eea{\end{eqnarray}}

\begin{document}
\vspace*{4cm}
\title{Recent Results from the ANTARES Neutrino Telescope}

\author{S. MANGANO, on behalf of the ANTARES Collaboration}

\address{IFIC - Instituto de F\'isica Corpuscular, Pol\'igono de la Coma,\\
46980 Paterna, Spain}

\maketitle\abstracts{
The ANTARES experiment is currently the largest underwater neutrino
telescope. It is taking high quality data since 2007 and aims to detect high
energy neutrinos that are expected from the acceleration of cosmic rays
from astrophysical sources. We will review the status of the detector
and present several analyses carried out on atmospheric muons and
neutrinos. For example we will show the latest results from searches
for neutrinos from steady cosmic point-like sources,
for neutrinos from Fermi Bubbles, for neutrinos from Dark Matter in the Sun and
the measurement of atmospheric neutrino oscillation parameters. 
}

\section{Introduction}
The discovery of high energy cosmic rays which 
collide with the Earth's atmosphere 
is known since 100 years, but their astrophysical 
origin and their acceleration to
such high energies is still unclear.
The observation of these cosmic rays is a strong argument 
for the existence of high energy neutrinos from the cosmos. 
Such cosmic neutrinos are expected to be emitted along with gamma-rays by
astrophysical sources in processes
involving the interaction of accelerated hadrons and the subsequent
production and decay of pions and kaons.
%The observation of cosmic neutrinos coming from distant sources 
%can provide valuable information about the 
%particle production mechanism in astrophysical sources. 
%
%The protons and high energy photons have a large probability to be absorbed 
%by the interstellar medium and the microwave background photons, which 
%implies that 
%the universe is not transparent for such cosmic particles. 

The advantage of using neutrinos with respect to cosmic particles like protons 
and photons is that they
are not deflected by magnetic fields and are weakly interacting.
The neutrinos point directly back to the source of emission and can provide
information about the source.

The disadvantage to detecting neutrinos 
is their extremely small interaction probability. A very large 
amount of matter is needed to have some interacting neutrinos. 
The flux of high energy cosmic neutrinos can be estimated 
from the observation of 
the rate of high energy cosmic rays. 
Theoretical calculations show that a detector of the size 
of about a cubic kilometer is needed to discover high energy cosmic neutrinos.

A cost effective way to detect high energies neutrinos is to use detector 
material found in nature, like water and ice.
%As the cost of a detector is proportional to its size, the cheapest 
%detector material is the one found in nature, water and ice.
The detector material has to be equipped by a three dimensional array of 
light sensors, so that muon neutrinos are
identified by the muons that are produced in charged current
interactions. These muons are detected by measuring the
Cerenkov light which they emit when charged particles move faster than 
the speed of light in the detector material.
The knowledge of the timing of the Cerenkov 
light recorded by 
the light sensors allows to reconstruct the trajectory of the muon 
and so to infer the arrival direction of the incident neutrino.
This technique is used in
large-scale Cerenkov detectors like IceCube~\cite{icecube} and ANTARES~\cite{Ant1} which are 
currently looking for high-energy
($\sim$TeV and beyond) cosmic neutrinos.

\section{ANTARES Neutrino Telescope}
The ANTARES detector is taking data since the first lines were deployed in 
2007. It is located in the
Mediterranean Sea, \mbox{40 km} off the French coast at \mbox{$42^{\circ} 50'$N, $6^{\circ} 10'$E}.
%The detection principle relies on the observation of
%Cerenkov light emitted by relativistic charged particles in water and 
%is optimized for the detection of muons energy of around 1 TeV.
The detector consists of twelve vertical
lines equipped with 885 photomultipliers (PMTs) in total, 
installed at a depth of about 2.5~km. 
The distance between adjacent lines is of the order of about 70 m.
Each line is equipped with up to 25 triplets of PMTs 
spaced vertically by 14.5 m. The
PMTs are oriented with their axis 
pointing downwards at $45^{\circ}$ from the vertical. 
The instrumented detector volume is about 0.02 $\textrm{km}^3$.
The design of ANTARES is optimized for the detection of upward going muons 
produced by neutrinos which have traversed the Earth, 
in order to limit the background from downward going atmospheric muons. 
The instantaneous field of view is $ 2 \pi\, \mathrm{sr}$ for neutrino 
energies between $10\ \mathrm{GeV}$ and $100\  \mathrm{TeV}$, 
due to selection of upgoing events.
Further details on the detector can be 
found elsewhere~\cite{Ant1}.

\section{Cosmic Point-Like Neutrino Sources}

\begin{figure}
 \setlength{\unitlength}{1cm}
 \centering
 \begin{picture}(18.5,6.5)
\put(-0.0,-0.4){\epsfig{figure=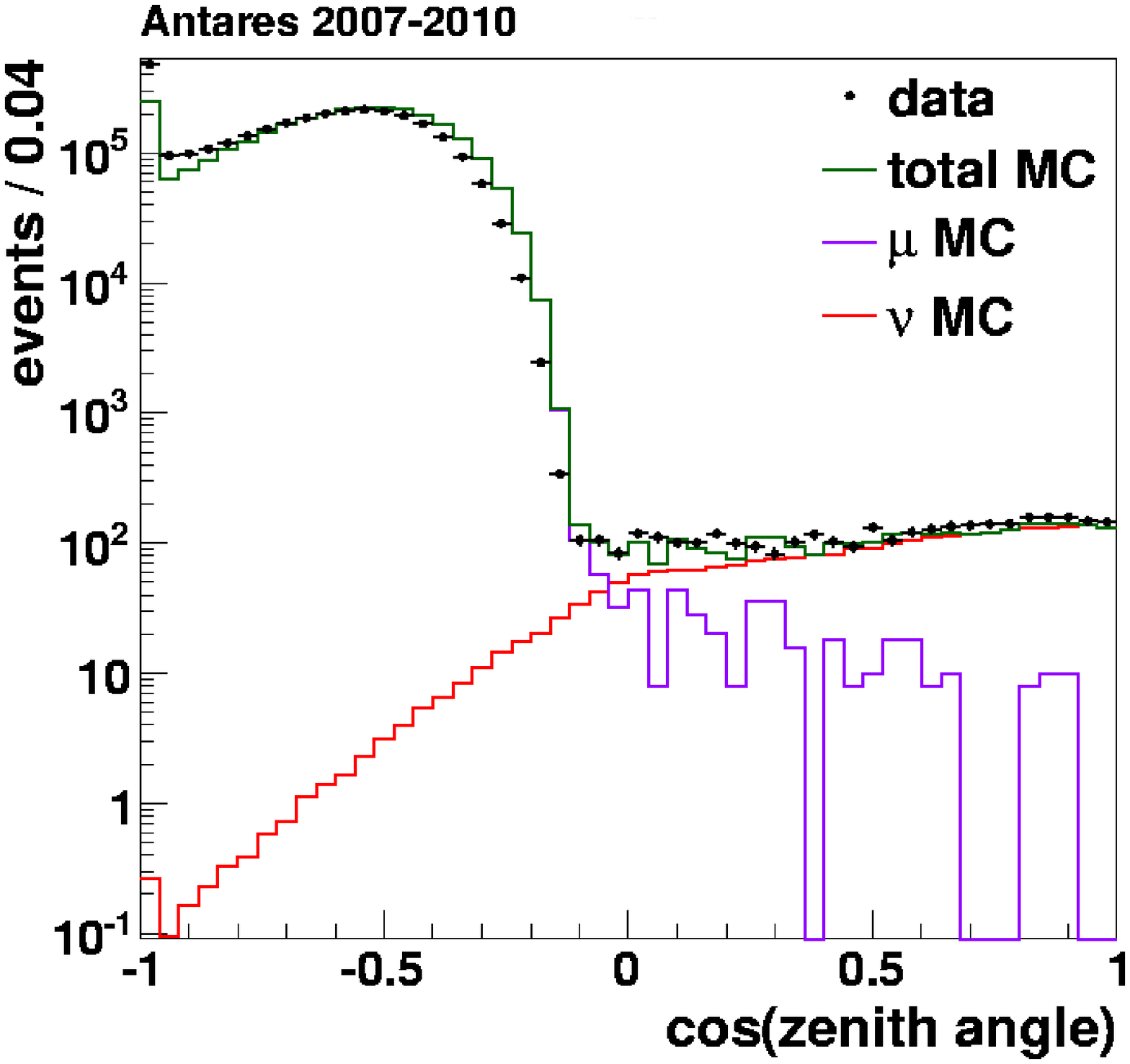,height=2.7in}}
\put(7.,0.0){\epsfig{figure=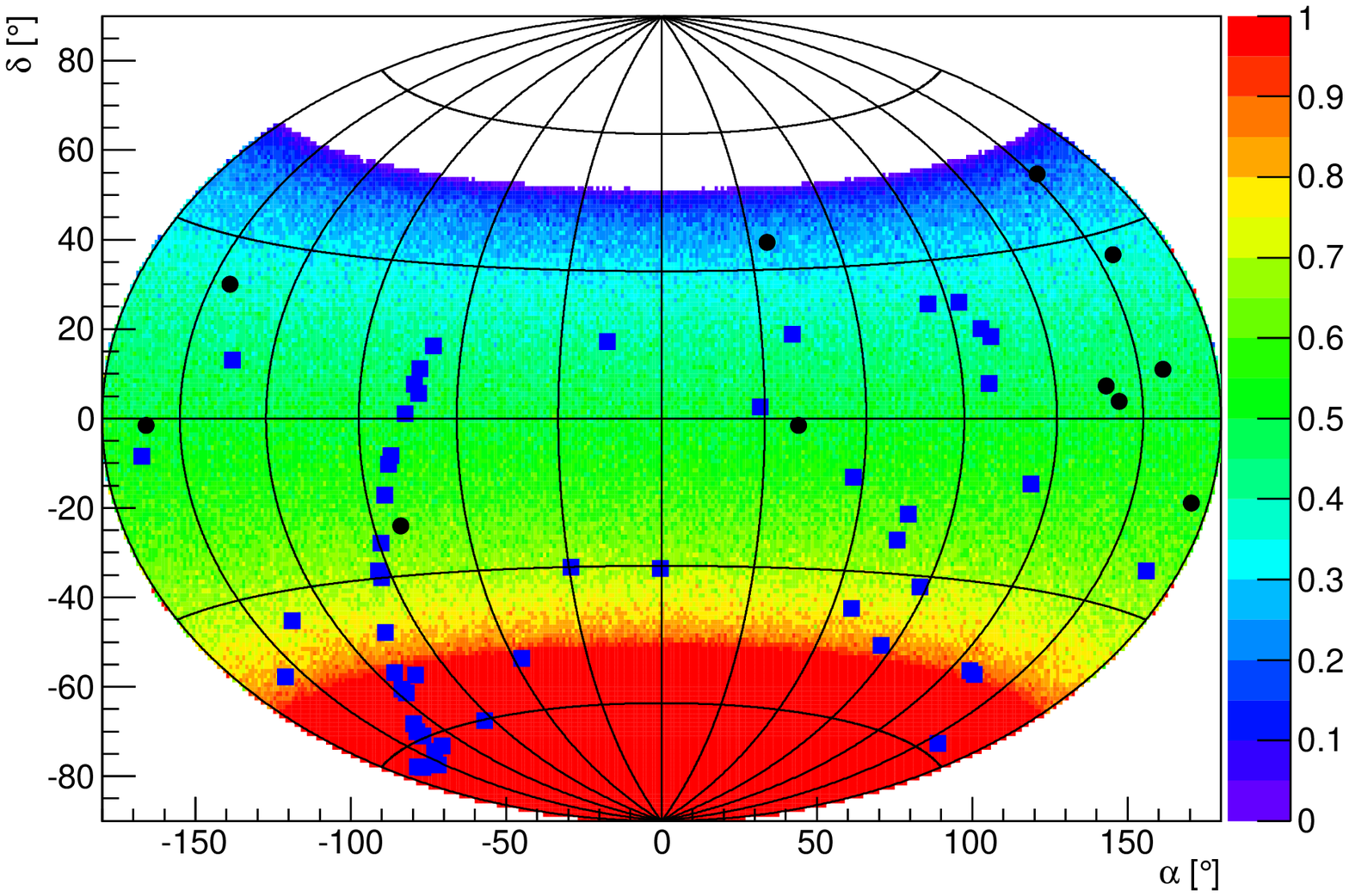,height=2.5in}}
\end{picture}
\caption {\textit {Left: The reconstructed  zenith angle data distribution of selected events compared to the Monte Carlo distribution for atmospheric neutrino and muon background. Right: Skymap in equatorial coordinates of the 51 neutrino candidate sources (blue squares) and the 11 most promising gravitational lensing systems (black circles). The ANTARES visibility is also shown.}}
\label{fig:zenithdiscovery}
\end{figure}

The main goal of the ANTARES neutrino telescope 
is the observation of neutrinos of cosmic
origin in the Southern sky. 
The main physical background to identify cosmic neutrinos are atmospheric muons
and upward going atmospheric neutrinos.
The atmospheric muons are produced in the upper
atmosphere by the interaction of cosmic rays and can 
reach the apparatus despite the
shielding provided by 2 km of water. 
Figure \ref{fig:zenithdiscovery} left shows a 
comparison of the zenith 
angle distribution between data and 
Monte Carlo simulation. It can be seen that 
the flux of atmospheric muons 
is several order of magnitude larger than that of atmospheric neutrinos
and that there is a good agreement between data and the Monte Carlo simulation.

The collaboration has 
developed several strategies to search in its data 
either for diffuse~\cite{diffusepaper}
or point-like~\cite{Pointsource2011,4yearpoint} cosmic neutrino sources, 
possibly in association with
other cosmic messengers such as gamma-rays~\cite{gammarayburst,gammaraypoint} 
or gravitational waves~\cite{gravitationalwave}.
Clustering of neutrino arrival directions 
can provide hints for their astrophysical origin.
In the search of cosmic neutrino point sources, upward going 
events have been selected in order to reject atmospheric muons. 
Additional cuts on the quality of the muon track reconstruction have 
been applied in order to eliminate events that correspond to downward going 
atmospheric muons which are misreconstructed as upward going. 
Most of the remaining events are atmospheric muon neutrinos 
which constitute an irreducible diffuse
background for cosmic neutrino searches.
The 2007-2010 data contain around 3000 neutrino candidates with a predicted
atmospheric muon neutrino purity of around 85\%. The estimated angular
resolution is 0.5$\pm$0.1 degrees.

The selection criteria are optimized 
to search for $E^{-2}$ neutrino flux from point-like 
astrophysical sources, following two different strategies: 
a full sky search and a search in the direction of 
particularly interesting neutrino candidate sources. 
The selection of 
these sources is either based on the intensity of 
their gamma-ray emission as observed by Fermi~\cite{Fermi} 
and HESS~\cite{Hess} or
based on strong gravitational lensed sources with large magnification.
%There are many examples of mass concentration 
%like galaxy clusters that enhance the photon
%flux observed from sources behind them through the gravitational
%lensing effect.
  
The motivation to select lensed sources is that 
neutrino fluxes as photon fluxes 
can be enhanced by the gravitational
lensing effect, which could allow to observe 
sources otherwise below the detection
threshold. Neutrinos (contrary to photons) are not 
absorbed by the gravitational lens. Therefore, sources with a moderate
observed gamma-ray flux could be interesting candidates for neutrino
telescopes. 
%The eleven most promising neutrino lensing systems have been selected 
%which have the potential to magnify any hypothetical cosmic neutrino flux
%into the observing capability of the ANTARES neutrino telescope.
%These candidates contain four lensed blazars and 
%five strongly amplified AGNs. In addition we also selected 
%the directions of two nearby and massive galaxy clusters with not
%known neutrino candidate source behind but which offer particularly
%magnification factors. 
Figure \ref{fig:zenithdiscovery} right shows the skymap in equatorial coordinates 
with the ANTARES visibility
of the 51 neutrino candidate sources with strong gamma-ray emission 
and the 11 most promising strong gravitationally lensed sources.

\begin{figure}
 \setlength{\unitlength}{1cm}
 \centering
 \begin{picture}(18.5,6.0)
\put(-0.,-0.0){\epsfig{figure=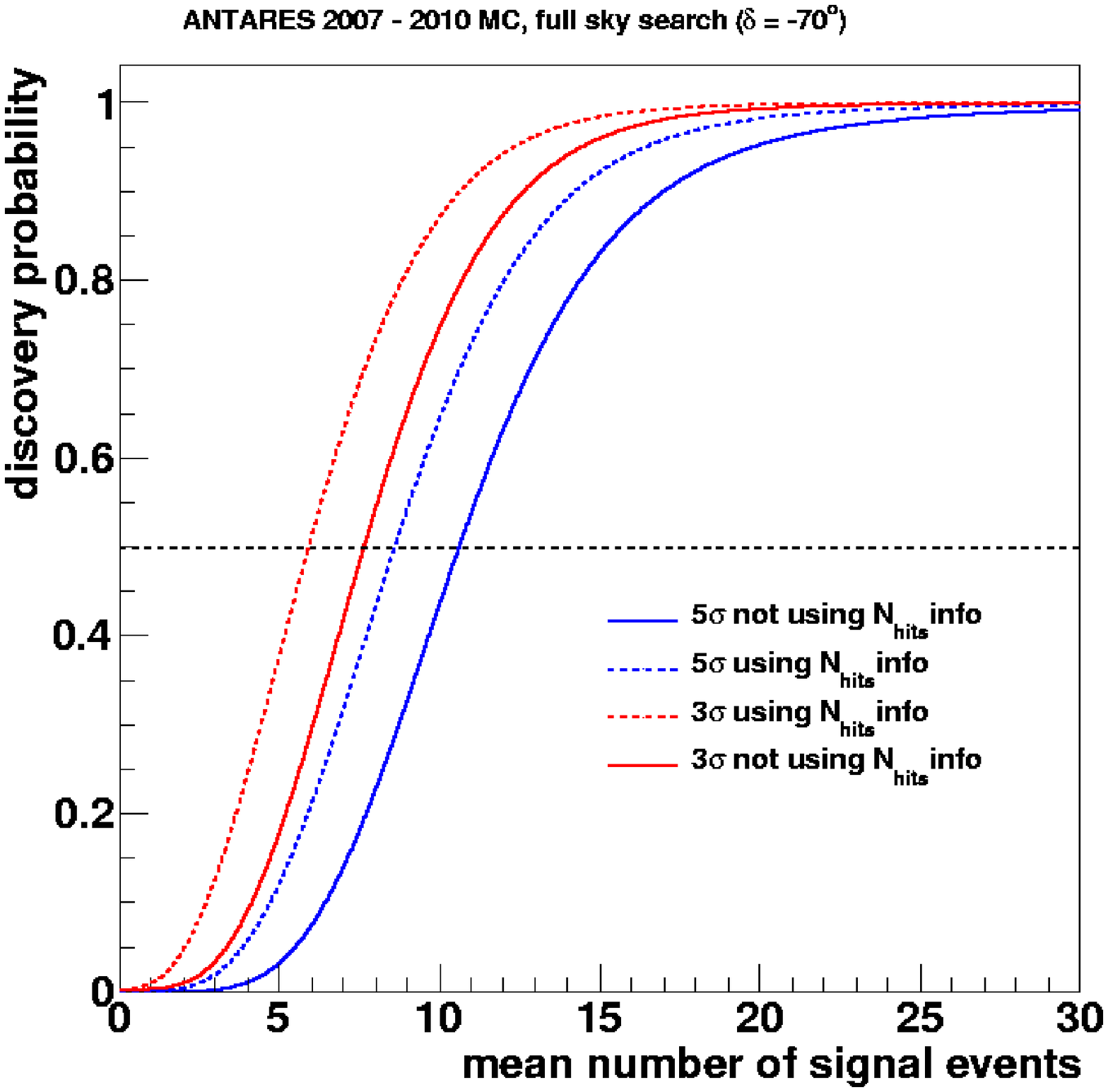,height=2.5in}}
\put(7.,-0.0){\epsfig{figure=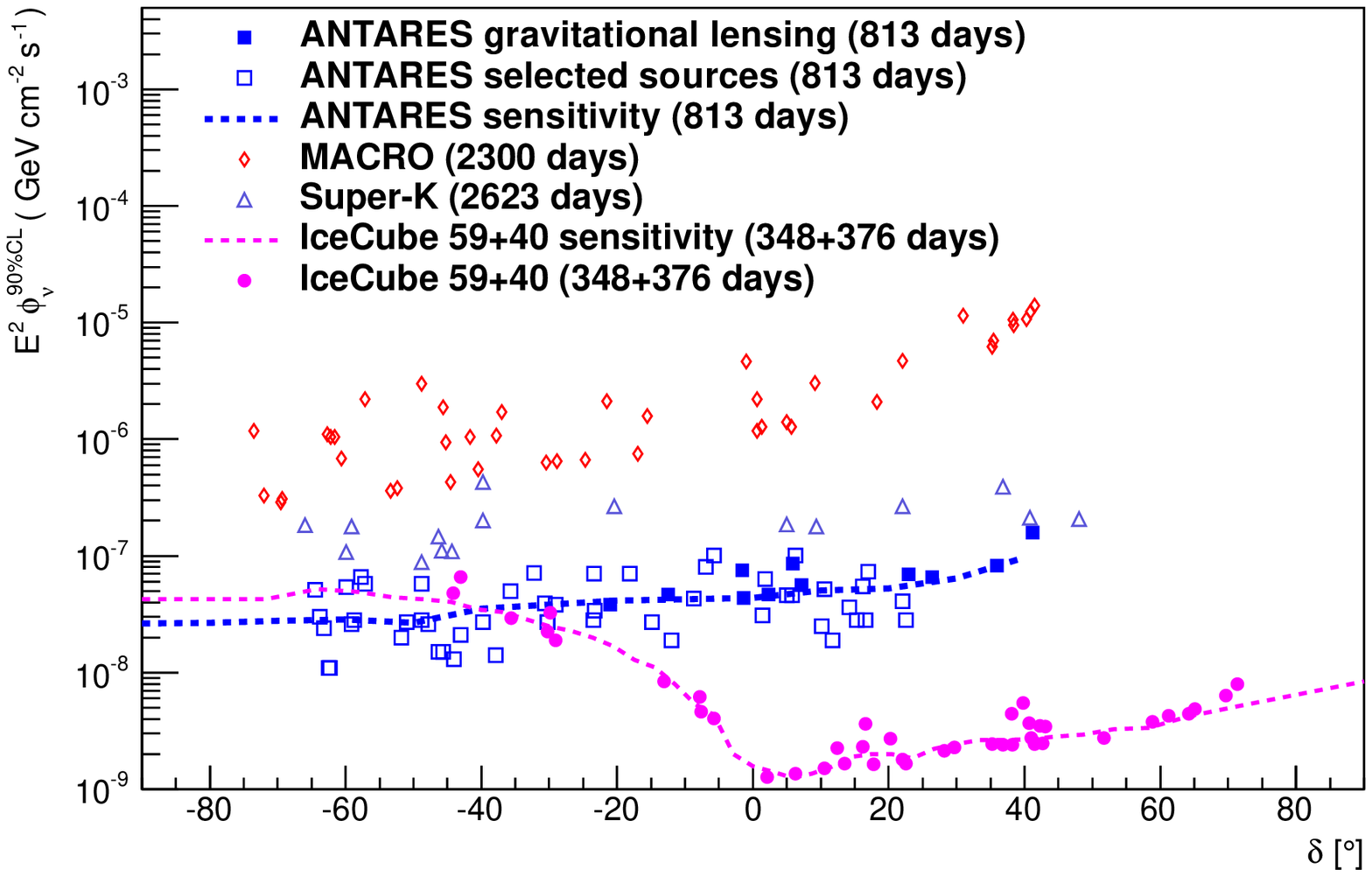,height=2.5in}}
 \end{picture}
\caption{\textit{Left: Probability for a 3 (red lines) and 5 (blue lines) sigma discovery as a function of the mean number of signal events. The dashed lines show the case where the number of hits information is used as an energy estimator in the likelihood, whereas for the solid lines no energy information is used. Right: Upper limits (at 90\% C.L.) on the $E^{-2}$ neutrino flux from the 62 selected candidate sources as well as the sensitivity as function of the declination. Also shown are the results from other experiments.}}
\label{fig:UpperLimits}
\end{figure}

The cosmic point source search has 
been performed using an unbinned maximum likelihood
method~\cite{4yearpoint}.  
This method uses
the information of the event direction and, since the cosmic sources are
expected to have a much harder spectra than atmospheric neutrinos,
the number of hits produced by the track. 
Figure \ref{fig:UpperLimits} left shows how the introduction of
an energy estimator like the number of hits increases
the discovery potential about 25\% in comparison 
without using such an information.
For each source, the position of the cluster is fixed at
the direction of the source and the likelihood function is maximized
with respect to the number of signal events. 
%After the
%likelihood maximization, a likelihood ratio is used as a test
%statistic.  The test statistic is the ratio between the
%value of the likelihood given by the maximization and the likelihood
%computed for the only-background case. Before unblinding, many
%pseudo-experiments are generated both for the case of only background
%and for the case of background with some signal added. After
%unblinding, the observed value is compared with the 
%distribution for the only-background case.
In the absence of a significant excess of
neutrinos above an expected background, an upper limit on the
neutrino flux is calculated.
%The flux upper limits are
%calculated at 90\% confidence level using the approach from
%Feldman \& Cousins~\cite{Feldman:1997qc}.

A full sky point source search based on the above mentioned algorithm
has not revealed a significant excess for any direction. 
The most significant cluster of events 
in the full sky search, with a post-trial 
$p$-value of 2.6\%, which is equivalent to $2.2\sigma$, corresponds to the 
location of $(\alpha,\delta)=(-46.5^{\circ},65.0^{\circ})$.  
No significant excess has been found also in
the dedicated search from the list
of 11 lensed and 51 gamma-rays 
selected  neutrino source candidates.
The obtained neutrino flux limits of 
these selected directions are plotted as function of declination in
\mbox{Figure \ref{fig:UpperLimits}} right,
where for comparison the limits set by other neutrino experiments
is also shown.

\section{Electromagnetic Showers along Muon Tracks}
Even if the primary aim of ANTARES is the detection of high energy
cosmic neutrinos, the detector measures mainly downward going muons. 
These muons are the decay products of cosmic ray 
collisions in the Earth's atmosphere.
Atmospheric muon data have been used for 
several analyses~\cite{4gev,anysotropy,fluxatmospheric}, 
in particular 
the collaboration investigated the sensitivity of 
the composition of cosmic rays through the
downward going muon flux~\cite{Hsu}. Several observational parameters
are combined to estimate the relative contribution of 
light and heavy cosmic rays. One of these parameters is the number of
electromagnetic showers along muon tracks.
 
Catastrophic energy losses appear occasionally, when a high energy muon 
($\sim 1$~TeV) traverses the water. 
These energy losses are characterized
by discrete bursts of Cerenkov light originating mostly from pair production
and bremsstrahlung (electromagnetic showers).
A shower identification algorithm~\cite{emshower,emshower1} 
is used to identify the excess of photons above 
the continuous baseline of photons emitted by
a minimum-ionizing muon. With this method downward going 
muons with energies up to 100 TeV have been analysed. 
%and count the 
%number of electromagnetic showers~\cite{emshower,salvi}
%per atmospheric muon event.
%The average muon energy loss per unit track length due to these 
%electromagnetic
%showers increases linearly with
%the energy of the muon allowing its energy to be determined.
%With this method downward going 
%muons with energies up to 100 TeV have been reconstructed. 
%to identify and 

\begin{figure}[tb]
 \setlength{\unitlength}{1cm}
 \centering
 \begin{picture}(18.5,6.0)
  \put(-0.,0.0){\epsfig{file=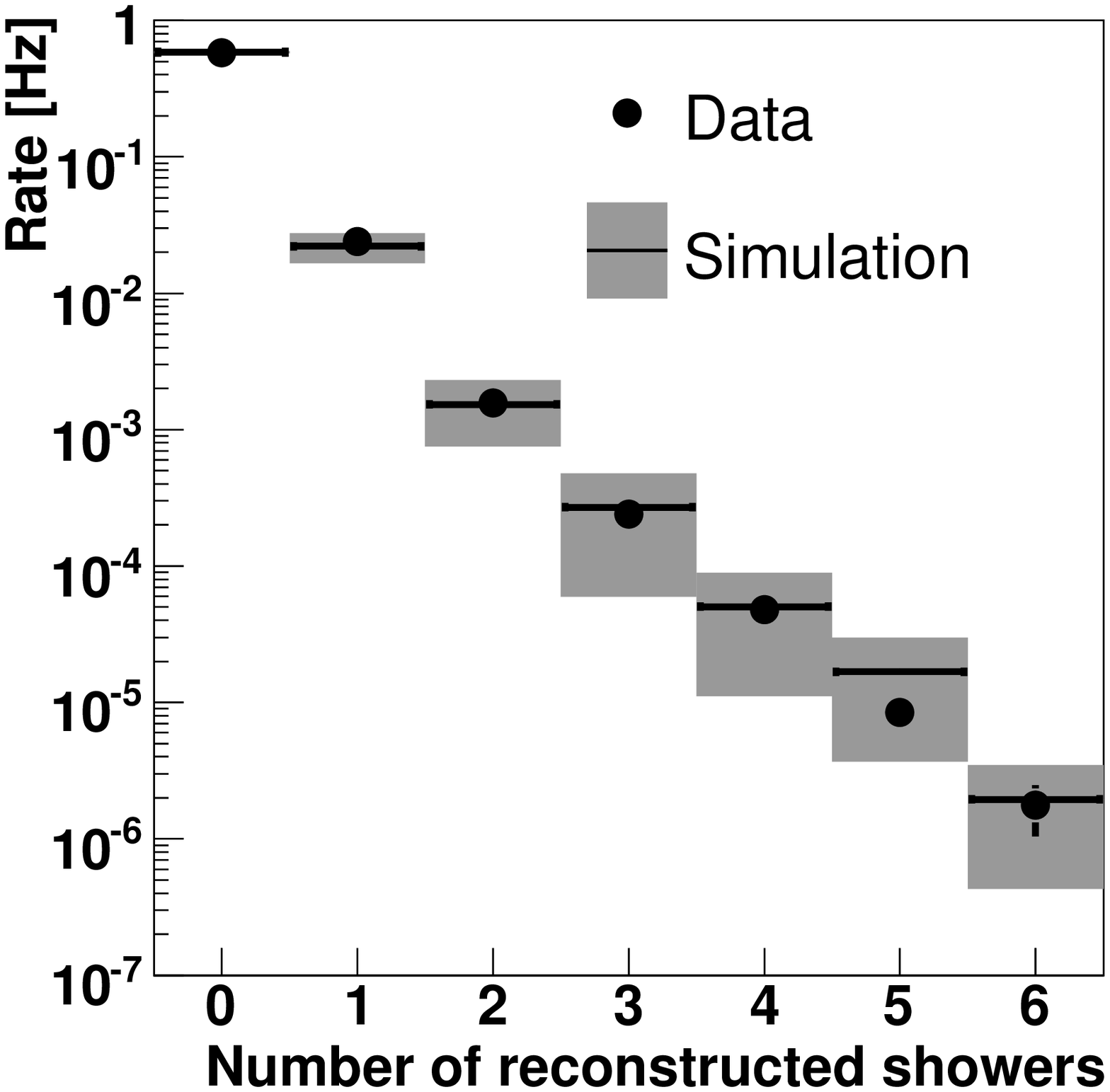,width=6.5cm,clip=}}
   \put(7., -0.1){\epsfig{figure=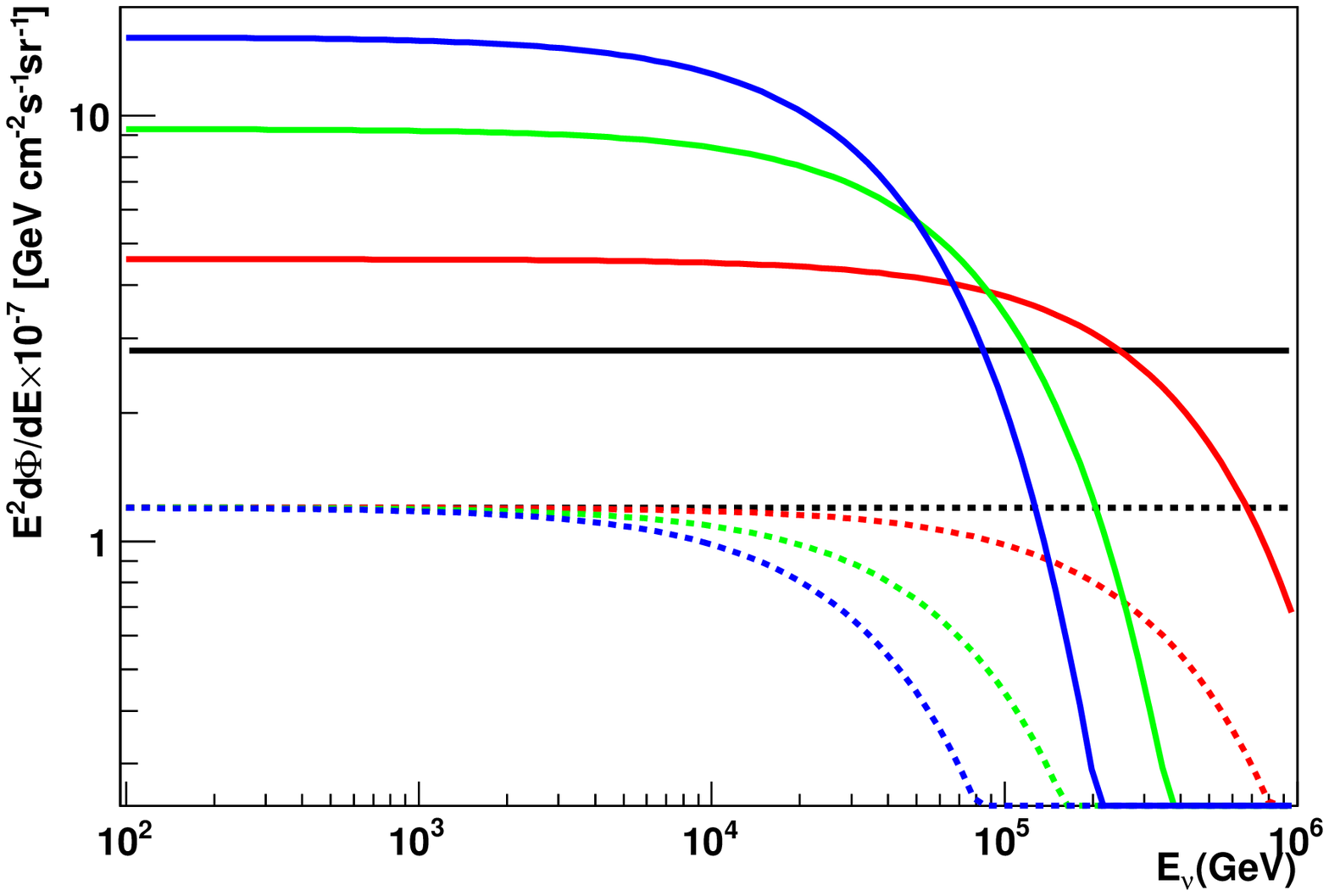,width=9.1cm,clip=}}
 \end{picture}
\caption[Sc]{\textit{Left: Muon event rate as a function of the shower multiplicity for data (points) and the Corsika simulation (line). The systematic error for the simulation is given by the height of the gray bands. Only statistical errors are shown for the data points. Right: Upper limits (at 90\% C.L.) on the $E^{-2}$ neutrino flux from the Fermi Bubbles for different cutoffs: no cutoff (black), 500 TeV (red), 100 TeV (green), 50 TeV (blue) are shown together with theoretical predictions in case of pure hadronic model (dashed lines).}}
\label{fig:showers}
\end{figure}

The muon event rate as a function of the number of identified
showers is plotted
in \mbox{Figure \ref{fig:showers}} left.
The distribution shows
the results for data and a Corsika based simulation.
As can be seen, about 5\% of the
selected muon tracks have at least one well identified shower.
Also shown is the systematic uncertainty for the simulation, 
where the largest systematic errors arises from uncertainties
on the PMT angular acceptance and absorption length.
%For the data points, only the statistical errors
%are presented.
%There is agreement between data and Monte Carlo over five orders of magnitude.

\section{Fermi Bubbles}
The Fermi Satellite has revealed an excess of gamma-rays
in an extended pair of bilateral bubbles above and below our Galactic Center.
These so called Fermi Bubbles (FB) cover about 0.8 sr of the sky,
have sharp edges, are relative constant in intensity and have a flat $E^{-2}$ 
spectrum between 1 and 100 GeV. 

It has been proposed that FB are seen due
to cosmic ray interactions with the interstellar medium, which produce pions. 
In this scenario gamma rays and high-energy neutrino emission are expected with
a similar flux from the pions decay.

ANTARES has an excellent visibility to the FB and therefore
a dedicated search for an excess of neutrinos in the
region of FB has been performed~\cite{FBubbles}. This analysis compares  
the rate of observed neutrino events in the region of the FB to 
that observed excluding the FB region. The of source FB
region is equivalent in size and has in
average the same detector efficiency as the FB region.
The analyzed 2008-2011 data reveal 16 neutrino events inside the
FB region. Estimations from outside the FB region predicts 
11 neutrino events. These results are compatible with no signal and
limits are placed on the fluxes of neutrinos 
for various assumptions on the energy 
cutoff at the source. Figure \ref{fig:showers} right shows the upper limits 
and compares it to expected signal 
for optimistic models~\cite{Fermib}. It can be seen 
that the calculated upper limits
are within a factor 3 above the expected signal.

\section{Dark Matter}

\begin{figure}
 \setlength{\unitlength}{1cm}
 \centering
 \begin{picture}(18.5,6.0)
\put(-0.2, 0.0){\epsfig{figure=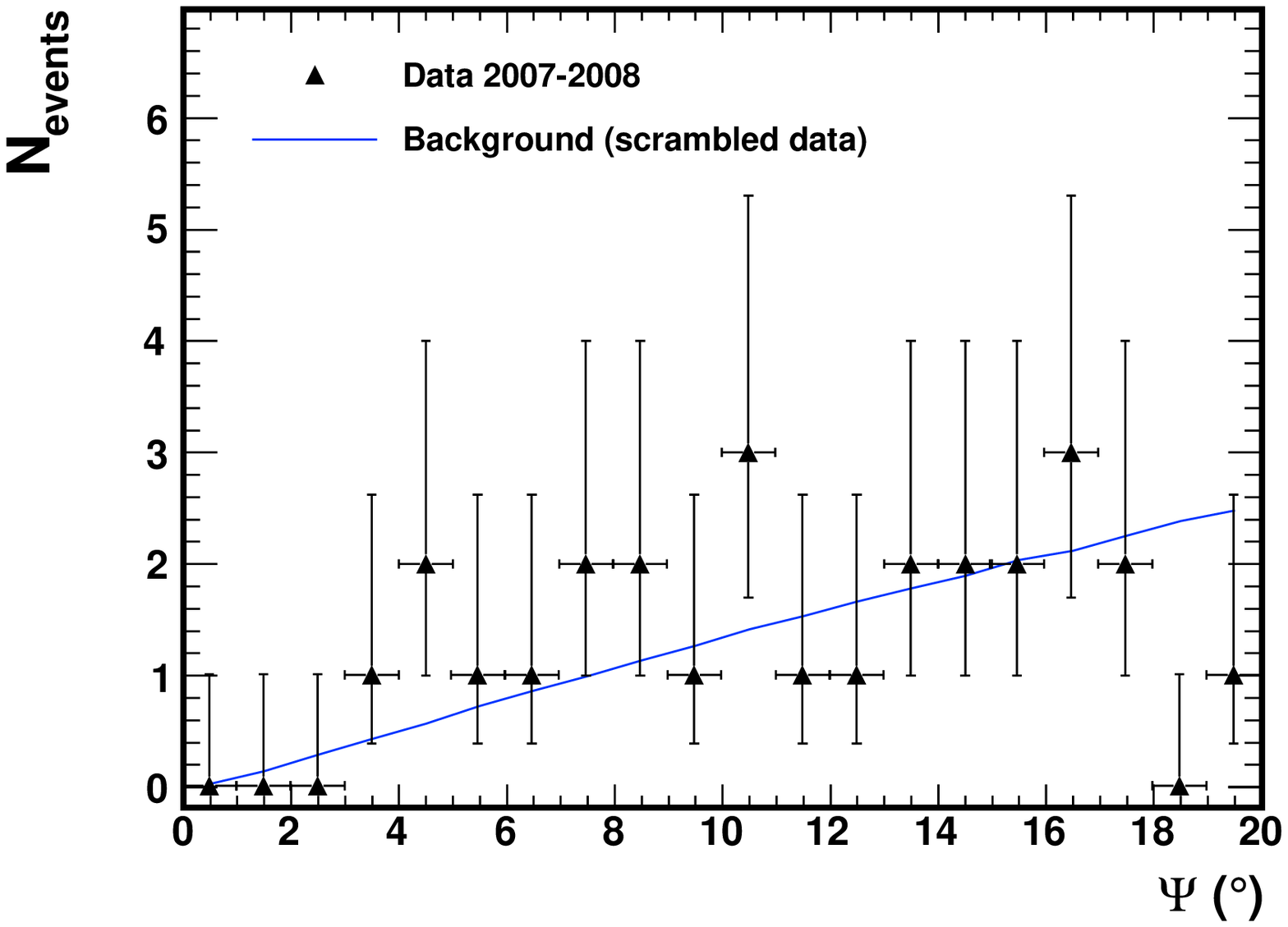,height=2.5in}}
\put( 7.8, 0.0){\epsfig{figure=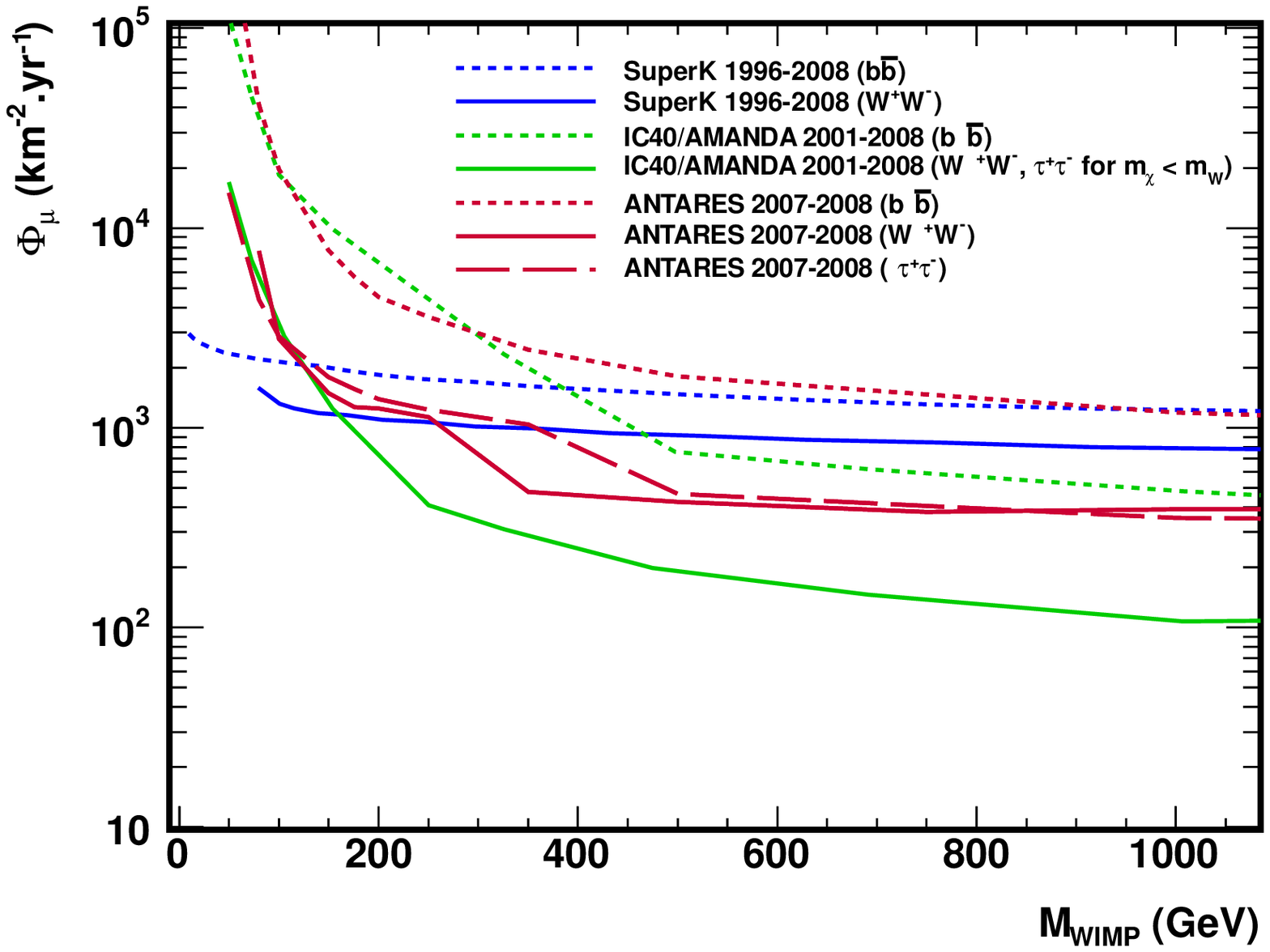,height=2.5in}}
\end{picture}
\caption{\textit{Left: The number of observed neutrinos (black triangles) and expected background events (solid blue line) as a function of the search cone radius around the Sun. A $1\sigma$  Poisson uncertainty is shown for each data point. Right: 90\% C.L. upper limit on the muon flux as a function of the WIMP mass for three self-annihilation channels $b\bar{b}$ (dotted line), $W^{+}W^{-}$ (solid line) and $\tau^{+} \tau^{-}$(dashed line). The results from Super-Kamiokande (in blue) and IceCube (in green) are also shown.}}
\label{fig:darkmatter}
\end{figure}

The indirect search for dark matter in the universe 
is one further goal of ANTARES~\cite{Dmatter}.
The weakly interactive massive particles could be gravitationally 
trapped in the center of the Sun. 
After the self-annihilation process
of these weakly interactive massive particles 
neutrinos can be created.

Using the data recorded during 2007 and 2008, a search for 
high energy neutrinos coming from the direction of the Sun has been performed. 
The neutrino selection criteria have been chosen to maximize 
a possible neutrino signal. 
Decay channels leading to both hard ($W^{+}W^{-}, \tau^{+} \tau^{-}$) and soft
($b \bar{b}$) flux spectra are considered.
The expected background is estimated
from the data by scrambling the direction of the observed neutrino candidates.
The number of events in a search cone around the Sun is shown in
Figure \ref{fig:darkmatter} left as a function of the search cone radius. 
It can be seen that the number of observed neutrino candidates 
is in good agreement 
with the background expectations. 
As there is no evidence for a flux of neutrinos from the Sun, upper limits 
on the muon flux are set and are presented 
in Figure \ref{fig:darkmatter} right. For comparison the limits from 
the other experiments such as  IceCube~\cite{icecubedm} and Super-Kamiokande~\cite{SuperK} are also shown.
By improving the event selection and including data already available
the analysis can be improved by around an order of magnitude.

\section{Velocity of Light in Water}

\begin{figure}
 \setlength{\unitlength}{1cm}
 \centering
 \begin{picture}(18.5,6.0)
\put(-0., 0.0){\epsfig{figure=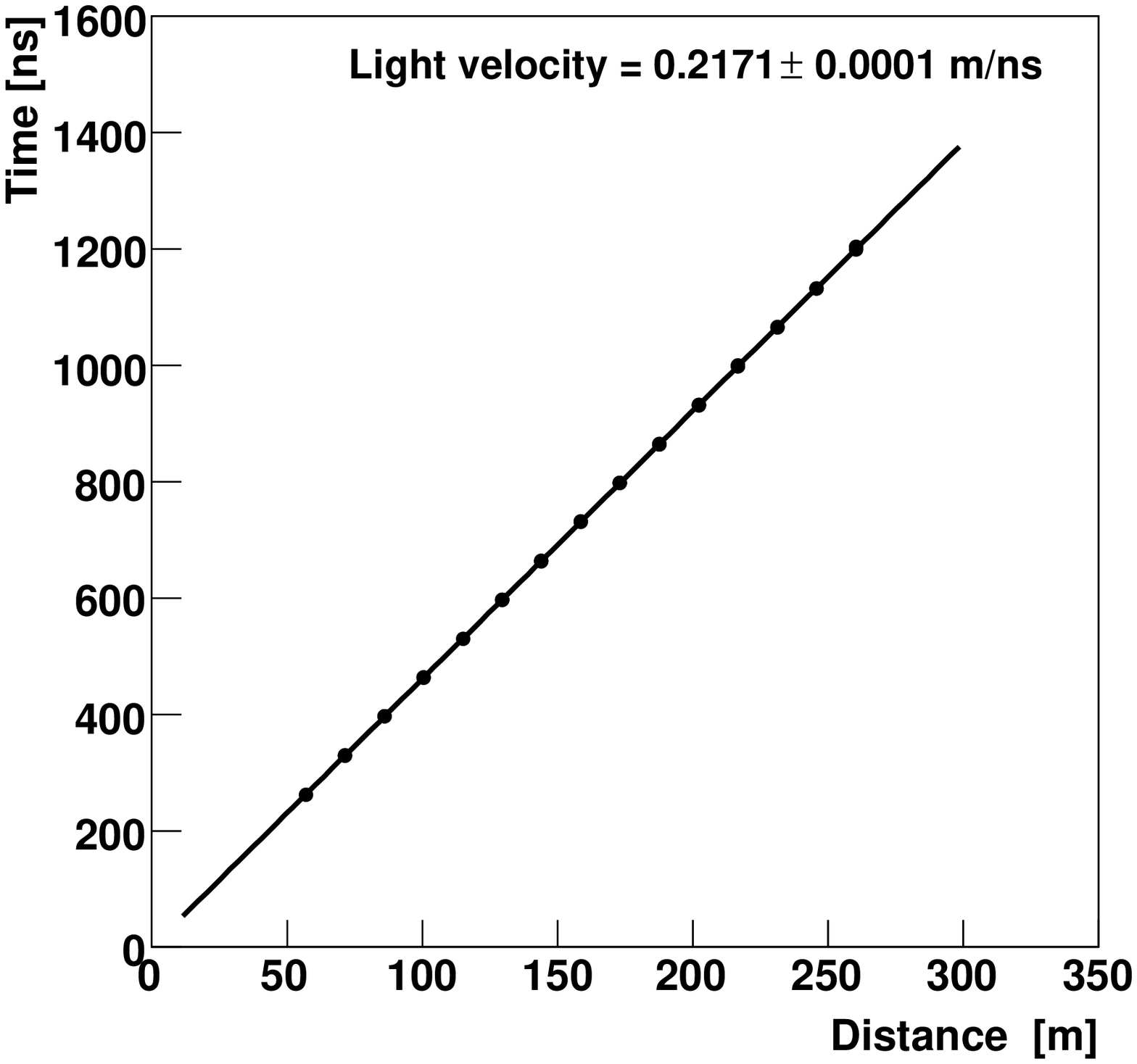,height=2.55in}}
\put(7.0, 0.0){\epsfig{figure=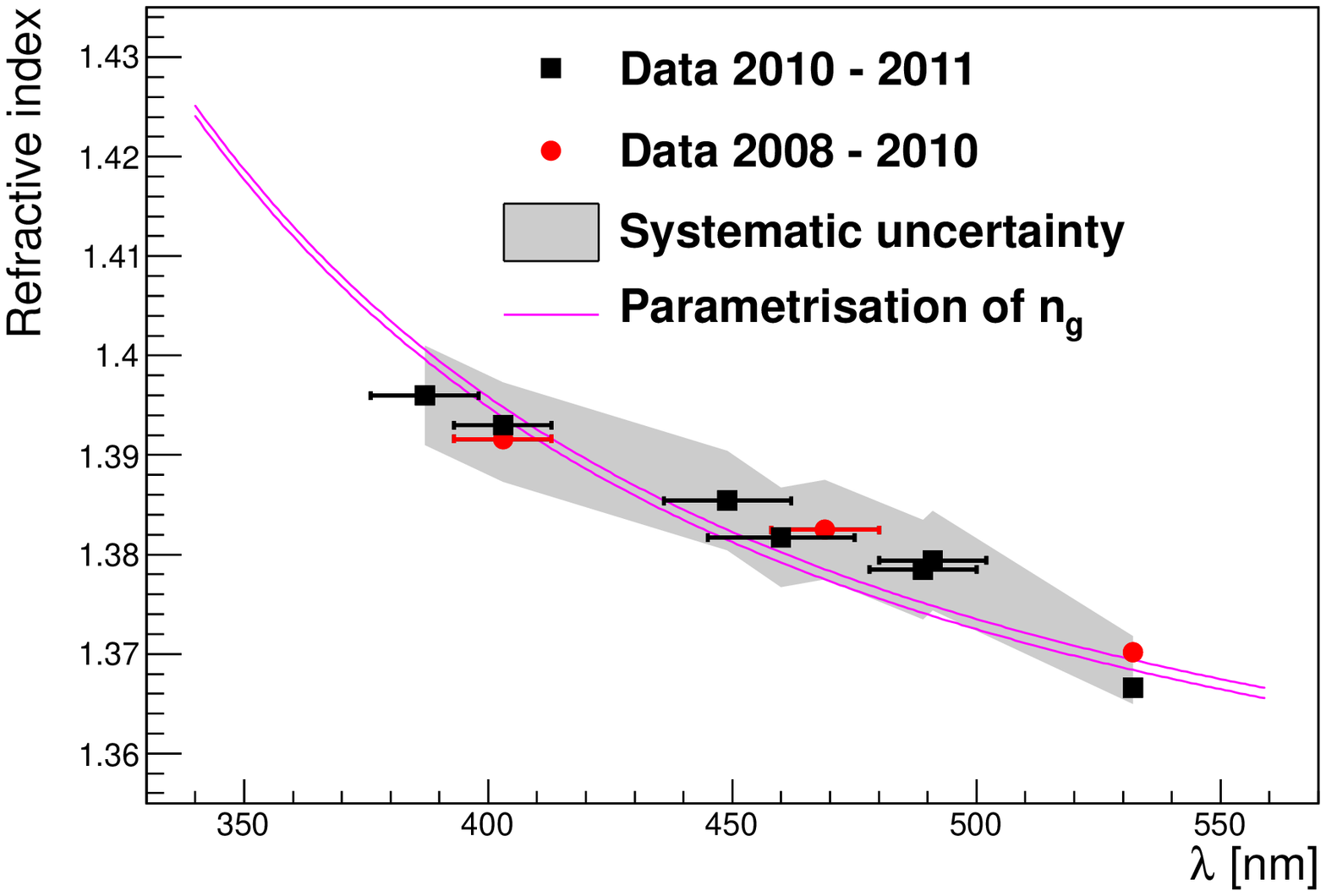,height=2.5in}}
 \end{picture}
\caption{\textit{Left: Arrival time as a function of the distance between the LED ($\lambda$ = 469~nm) and the PMT. The solid line corresponds to a fit of a linear function to the data. Right: Index of refraction corresponding to the group velocity of light as a function of the wavelength. The gray band shows the systematic uncertainty. The two solid lines correspond to a parametrization of the index of refraction evaluated at a pressure of 200 atm (lower line) and 240 atm (upper line).}} 
\label{fig:velocityoflight}
\end{figure}

%The propagation of Cerenkov
%light depends on the optical properties of the sea water
%and their understanding is crucial to reach the optimal performance
%of the detector. 

Charged particles crossing sea water induce the emission of Cerenkov
light whenever the condition $\beta > 1/n_p$ is fulfilled, where
$\beta$ is the speed of the particle relative to the
speed of light in vacuum and $n_p$ is the phase refractive index. 
The Cerenkov photons 
are emitted at an angle with respect
to the particle track given by $\cos \, \theta_c = \frac{1}{\beta n_p}$.
The individual photons then travel in the medium at the group velocity. 
Both the phase and group refractive indices depend on the
wavelength of the photons and has the effect of making the emission
angle and the speed of light wavelength dependent. Good knowledge of this wavelength 
dependence enables to reach the optimal performance of the detector.

The velocity of light has been measured using
a set of pulsed light sources
(LEDs emitting at different wavelengths) 
distributed throughout the detector illuminating the
PMTs through the water~\cite{velocity}.  
In special calibration data runs the emission
time and the position of the isotropic light flash,
as well as the arrival time and the position when the light reaches 
the PMTs are recorded.
Figure \ref{fig:velocityoflight} left shows the arrival time as a 
function of the distance between the LED and the different PMTs. 
The slope of a linear fit to the arrival time versus distance 
gives the inverse of the light velocity.
The refractive index has been measured 
at eight different wavelengths between 385~nm and 532~nm.  
This refractive index with its 
systematic errors are shown in Figure \ref{fig:velocityoflight} right.
Also shown is the parametric formula of the refractive index. 
The measurements are in agreement with the 
parametrization of the group refractive index.

\section{Atmospheric Neutrino Oscillations}
ANTARES is also sensitive to neutrino oscillation parameters 
through the disappearance of atmospheric muon neutrinos~\cite{oscillation}.

Neutrino oscillation are commonly described
in terms of L/E, where L the oscillation path length 
and E is the neutrino energy. 
For upward going neutrinos crossing the Earth  
the travel distance L is translated to $D \cos\theta$ where
$D$ is the Earth diameter and $\theta$ the zenith angle.
Within the two-flavour approximation, the $\nu_{\mu}$
survival probability can be written as

$$P(\nu_{\mu} \rightarrow \nu_{\mu}) = 1 - \sin^2 2\theta_{23} \cdot \sin^2(1.27 \triangle m^2_{23}\frac{L}{E_{\nu}}) = 1 - \sin^2 2\theta_{23} \cdot \sin^2(16200 \triangle m^2_{23}\frac{\cos \theta}{E_{\nu}}),$$

where $\theta_{23}$ is the mixing angle and $\triangle m^2_{23}$ is the squared
mass difference of the mass eigenstates (with $L$ in km, $E _{\nu}$ in GeV and $\triangle m^2_{23}$ in eV$^2$). The survival probability $P$ depends
only on the two oscillation parameters, $sin^2 2\theta_{23}$ and $\triangle m^2_{23}$,
which determine the behavior for the atmospheric neutrino oscillations.

Taking the recent results from the MINOS experiment~\cite{minos}, the first minimum
in the muon neutrino survival probability 
$(P(\nu_{\mu} \rightarrow \nu_{\mu})=0)$ occurs for vertical upward going neutrinos
at about 24 GeV. 
Muons induced by a 24 GeV neutrino travel in average around 120 m in sea water.
The detector has PMTs spaced vertically by 14.5 m so that this energy range 
can be reached for events detected on one single line. 
%and a clean sample of atmospheric neutrinos 
%with energies as low as 20 GeV are isolated.

\begin{figure}
 \setlength{\unitlength}{1cm}
 \centering
 \begin{picture}(18.5,6.5)
\put(-0.1, 0.0){\epsfig{figure=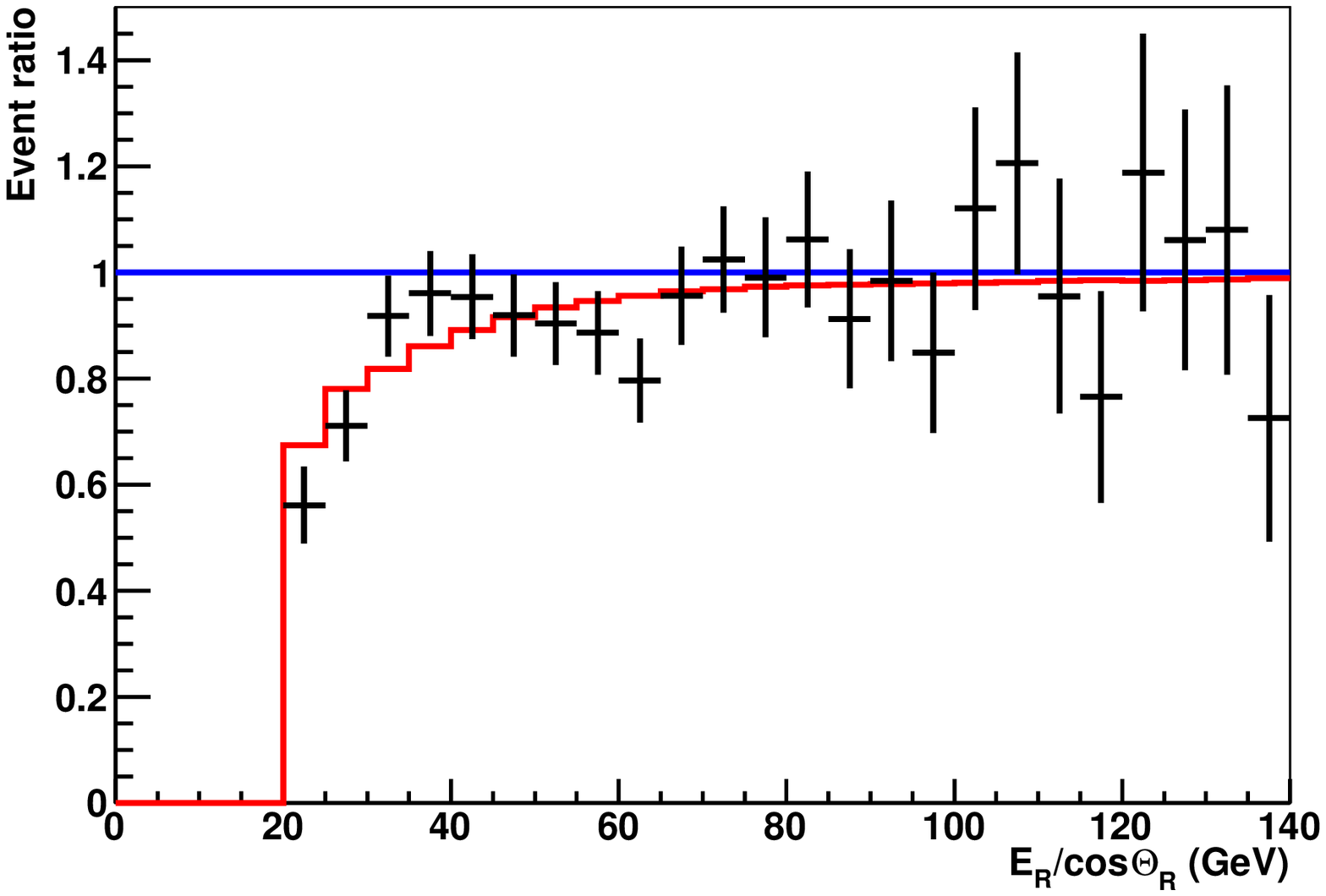,height=2.3in}}
\put( 8.0, 0.0){\epsfig{figure=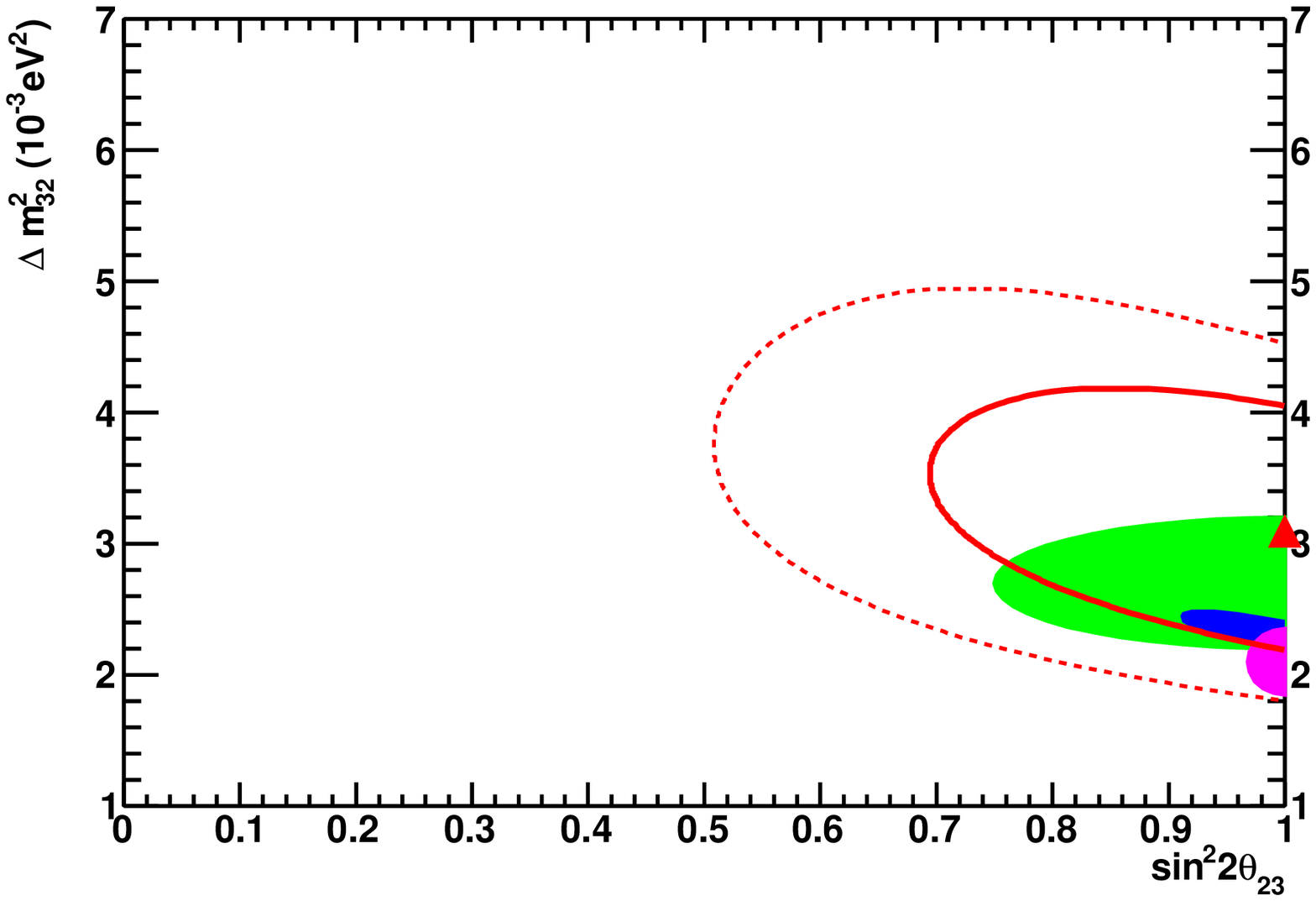,height=2.3in}}
\end{picture}
\caption{\textit{Left: The fraction of events as a function of the $E_R/cos\theta_R$ distribution. Black crosses are data with statistical uncertainties, the blue histogram shows the non-oscillation hypothesis and the red histogram shows the result of the best fit. Right: 68\% and 90\% C.L. contours (solid and dashed red lines) of the neutrino oscillation parameters as derived from the fit of the $E_R/cos\theta_R$ distribution. The best fit point is indicated by the triangle. For comparison the solid filled regions show results at 68\% C.L. from K2K (green), MINOS (blue) and Super-Kamiokande (magenta).}}
\label{fig:oscillation}
\end{figure}

The reconstructed flight 
path through the Earth is reconstructed through zenith angle $\theta_R$, 
which is estimated from a muon track fit~\cite{fastalgo}. Whereas
the neutrino energy $E_R$ is estimated from 
the observed muon range in the detector.
Figure \ref{fig:oscillation} left 
shows event rate of the the measured variable 
$E_R/cos\theta_R$ for a data sample from 2007 to 2010 with 
a total live time of 863 days.
The neutrino oscillations causes a clear 
event suppression for $E_R/cos\theta_R < 60 GeV$ with a clean sample of 
atmospheric neutrinos 
with energies as low as 20 GeV.
The parameters of the atmospheric neutrino oscillations 
are extracted by fitting the event rate as a function $E_R/cos\theta_R$ and
is plotted as a red curve in Figure \ref{fig:oscillation} left with
values $\triangle m^2_{23}=3.1\cdot 10^{-3} eV^2$ and $sin^2 2\theta_{23} = 1$.

This measurement is converted into limits of the oscillation parameters and
is shown in Figure \ref{fig:oscillation} right. 
If maximum mixing is imposed ($sin^2 2\theta_{23} = 1$) the values of 
$\triangle m^2_{23}$ is 
$\triangle m^2_{23}=(3.1 \pm 0.9) \cdot 10^{-3} eV^2$.
This measurement is in good agreement with the world average measurements.
Although the results are not competitive with dedicated experiments,
the ANTARES detector demonstrates the capability to measure
atmospheric neutrino oscillation parameters 
and to detect and measure energies as low as 20 GeV.

\section{Conclusion}
ANTARES has been taking data since the first lines were deployed in 2007.
With these data a broad physics program is underway 
producing competitive results. Unfortunately ANTARES has still not seen 
any cosmic neutrinos.   

In this proceeding there was not 
enough room to discuss other
topics which were illustrated at the Rencontres de Moriond, 
such as the atmospheric muon~\cite{4gev}
 and neutrino fluxes~\cite{diffusepaper}, the time \mbox{calibration} system~\cite{timecalib}, the acoustic neutrino detection system~\cite{acoustic}, the search for relativistic magnetic monopoles~\cite{monopol}, the searches for nuclearites~\cite{nuclearites} and the correlation of neutrinos with gravitational waves~\cite{gravitationalwave}, for which
the reader is referred to elsewhere.

Neutrino telescopes 
are starting to open up a new window in the sky exploring new territory
and they will hopefully reveal new unknown phenomena and 
help answer open questions.

\section*{Acknowledgments}
\vspace*{-0.1cm}
I gratefully acknowledge the support of the JAE-Doc postdoctoral program of CSIC.
This work has also been supported by the following
Spanish projects: FPA2009-13983-C02-01, MultiDark Consolider CSD2009-00064, ACI2009-1020 of MICINN and Prometeo/2009/026 of Generalitat Valenciana.

\end{document}